\documentclass[prd,aps,tightenlines,superscriptaddress,11pt]{revtex4}
\usepackage{epsfig}
\begin{document}
\title{$\Delta\rightarrow N\gamma^*$ Coulomb Quadrupole Amplitude in pQCD}
\author{Ahmad Idilbi} \email{idilbi@physics.umd.edu}
\affiliation{Department of physics,University of
Maryland,College-Park,Maryland 20742, USA}
\author{Xiangdong Ji}
\email{xji@physics.umd.edu} \affiliation {Department of
physics,University of Maryland,College-Park,Maryland 20742, USA}
\author{Jian-Ping Ma}
\email{majp@itp.ac.cn} \affiliation{Institute of Theoretical
Physics, Academia Sinica, Beijing, 100080, P. R. China}
\begin{abstract}

We present a leading-order pQCD calculation of the helicity-flip $\Delta\rightarrow N\gamma^*$ matrix element
$G_0$ (Coulomb quadrupole amplitude $C2$), taking into account the transverse momenta of the quarks and the
contribution from the gluons. In the large $Q^2$ limit, its scaling behavior acquires a double-logarithmic
correction $\log^2{(Q^2/\Lambda^2)}$ compared with the standard scaling analysis, due to the contribution from the
orbital motion of the small-$x$ partons. Based on this and on the latest JLab experimental results of the $C2-M1$
ratio $R_{SM}$ at $Q^2$ = 3 $\sim$ 4 GeV$^2$, we make a phenomenological prediction for the latter at higher
values of $Q^2$.
\end{abstract}
\maketitle

\section{introduction}

In recent years, there have been continuing interests in the electromagnetic $N\rightarrow \Delta$ transition. One
of the earlier interests was due to the work of Becchi and Morpurgo \cite{becc}. In that work they had shown that
in the context of the symmetric, non-relativistic, $SU(6)$ quark model, the transition $N \rightarrow \Delta$ is
a pure magnetic dipole $M1$ and the contribution from the electric quadrupole $E2$ is zero. One crucial
assumption in their derivation of this `selection rule' is that the quarks in both the nucleon and the delta are
in the zero orbital angular momentum states.

These predictions were to be considered as a check to the
validity of the quark model, which was still questionable in those
days. Later experimental measurements \cite{dal} showed that,
indeed, the magnetic dipole $M1$ contributes predominantly to the
$N \rightarrow \Delta$ transition, while the contribution from
$E2$ is small but non-vanishing.

The non-vanishing value of $E2$, and also the Coulomb quadrupole
$C2$ in the case of a virtual photon,  has generated much
theoretical interest. One way to account for $E2 \neq 0$ is
through the $D$-wave mixtures in the $N$ and $\Delta$ wave
functions \cite{isgure}. Another way is through the two-body
electromagnetic currents from one-gluon and/or one-pion exchange
between constituent quarks \cite{buch}. In the latter case, it is
argued that $E2$ transition is, mainly, due to a two-quark
spin-flip operator. On the other hand, in the large $N_c$ limit of
quantum chromodynamics it has been shown \cite{jenk} that $R_{EM}
\equiv E2/M1 $ is of order $1/N_c^2$. To derive this result, no
assumption about orbital angular momentum of the quarks was
necessary. More recent work in this direction can be found in
Ref. \cite{buchi}.

Another major issue related to the $N \rightarrow \Delta$
transition, and in general to any hadronic exclusive process, is
the applicability of perturbative QCD (pQCD) at the range of
values of momentum transfer $Q^2$ accessible in the current
generation of experiments. In terms of the ratios $R_{EM}$ and
$R_{SM} \equiv C2/M1$,\ pQCD power counting predicts that
\cite{carl}, in the limit $Q^2 \rightarrow \infty$, $R_{EM}
\rightarrow 1$ and $R_{SM} \rightarrow {\rm const.}$, up to
logarithmic corrections to be discussed in this paper. The former
prediction has  not yet been observed experimentally. In fact, up
to $Q^2=4$ GeV$^2$, $R_{EM}$ stays negative and very close to
zero \cite{beck, kall, fro, yang, lee}. The comparison between
data and pQCD prediction for $R_{SM}$ is the main topic of this
paper.


To make the pQCD predictions more relevant at finite $Q^2$ where data have been and will be taken, one has to go
beyond the asymptotic power counting, make detailed pQCD calculations of hard scattering amplitudes and derive the
factorization formula for the experimental observables. In the present case, the relevant quantities are the
three independent helicity matrix elements \cite{carl}:
\begin{equation}
G_\lambda =\frac{1}{2M_N} \langle\Delta,\lambda_\Delta \vert
\varepsilon(\lambda) \cdot J\vert P,\lambda_P=1/2 \rangle \ ,
\end{equation}
where $M_N$ is the mass of the nucleon,  $\lambda = 0 , \pm 1$,
and $\lambda_\Delta + 1/2=\lambda$. A pQCD calculation of the
helicity-conserving matrix element $G_+$ has been known for many
years \cite{brod1, lep, carl}. However, a pQCD calculation of the
helicity non-conserving amplitudes has not been explored in the
literature until recently, because {\it it necessarily involves
the orbital motion of the quarks} \cite{ji1,beli}.

In this paper, we perform a pQCD calculation for the
helicity-flip matrix element $G_0$. The technical details are
basically similar to those in the calculation of the Pauli form
factor $F_2(Q^2)$ of the proton performed in \cite{beli}. In the
calculation, the quarks in the nucleon and the $\Delta$ have one
unit of relative orbital angular momentum along the direction of
motion (taken as the $z$ axis) {\it i.e.}, $\vert \Delta l_z
\vert=1$. With this we calculate the ``hard amplitudes'' which,
in turn, are to be convoluted with the soft light-cone
distribution amplitudes.  The light-cone wave functions of the
nucleon and delta have been classified according to their orbital
angular momentum dependence in \cite{ji1,ma}. Note that we have to
add the gluon contribution  to maintain color gauge invariance.
We  neglect the dynamical gluon effects proportional to the gluon
field strength $F^{\alpha\beta}$ which seems numerically
suppressed in general \cite{ww}.

We remark that a detailed calculation of the hard amplitudes for $G_-$ will involve two units of quark relative
orbital angular momentum. Due to computational complexity it will not be pursued here. Nevertheless, it can be
easily shown that $G_-$ is suppressed, in the high $Q^2$ limit by $O(1/Q)$ relative to $G_0$ \cite{carl,ji1}.

Assuming our calculation for $G_0$ is relevant at $Q^2$ that is
currently explored at Jefferson Lab, our result and a recent
experimentally measured values of $R_{SM}$ enable us to make a
phenomenological prediction of this ratio for higher values of
$Q^2$. We mention in this regard that the double logarithmic term,
$\log^2(Q^2/\Lambda^2)$, will play an important role in our
analysis. The importance of $\log^2(Q^2/\Lambda^2)$ has also been
demonstrated recently in the Jefferson Lab data on the nucleon
elastic form factor $Q^2F_2/F_1$ \cite{beli}.

Our plan of representation is as follows: in section II we give
the notations and definitions needed for the coming sections. In
section III we present in some detail the calculations performed
and  comment on the obtained results. In section IV we perform a
phenomenological analysis  and make a prediction for $R_{SM}$ .
Section V will serve as a summary .

\vfill

\section{Kinematics and Notation}

In this paper, we treat the bare delta as if it were a bound state of QCD, although in experiment it appears only
as a resonance in certain scattering cross sections. Our analysis is relevant for scattering at the resonance peak
where the contribution from the background vanishes by definition. We will not consider the so-called the
dressing of the resonance due to the pion cloud which might be important at small $Q^2$ \cite{yang,lee}.

Consider the following scattering:
\begin{equation}
  P(P_N) +\gamma^* (q) \to \Delta (P_\Delta).
\end{equation}
which appears as a sub-process of the electro-production of pions.
We are interested in the $G_0$ matrix element, in which case the
photon is logitudinally-polarized and with a large virtuality
$Q^2=-q^2$. Our analysis will be carried out in a frame in which
the virtual photon $\gamma^*$, the incoming proton and the
outgoing $\Delta$ are collinear. The three momenta of the proton
and the $\Delta$ are in  opposite directions and the $\gamma^*$
is moving in the $-z$ direction. The photon polarization vector
is given by:
\begin{equation}
 \varepsilon(\lambda= 0) = \frac{-1}{\sqrt{Q^2}}(q^3,0,0,q^0).
\end{equation}
with $q^\mu=(q^0,0,0,q^3)$.

We will also need the light-cone wave functions of the proton and
the $\Delta^+$ (with the same charge as the proton). These wave
functions are given by \cite{ji1,ma}:
\begin{eqnarray}
\vert P,\lambda=1/2 \rangle&=& \frac{1}{12} \int
d[1][2][3]\varepsilon^{ijk}\left\{\psi_P^{(1)}(1,2,3)
 u_{i\uparrow}^\dagger(1)\left[
u_{j\downarrow}^\dagger (2)d_{k\uparrow}^\dagger
(3)-d_{j\downarrow}^\dagger(2) u_{k\uparrow}^\dagger (3)\right]
|0\rangle \right.\nonumber \\
&&+ \left(k_{1\perp}^+\psi_P^{(3)}(1,2,3)+k_{2\perp}^+
\psi_P^{(4)}(1,2,3)\right)\nonumber
\\
&& \left.\times \left[u_{i\uparrow}^\dagger
(1)u_{j\downarrow}^\dagger (2)d_{k\downarrow}^\dagger
(3)-d_{i\uparrow}^\dagger(1) u_{j\downarrow}^\dagger
(2)u_{k\downarrow}^\dagger(3)\right]|0\rangle\right\} + ... \ ,
\end{eqnarray}
\begin{eqnarray}
\vert \Delta^+,\lambda=-1/2 \rangle &=& - \frac{1}{12} \int d [1'][2'][3']\varepsilon^{ijk}
\left\{ \psi_{\Delta}^{(1)} (1',2',3') \right. \nonumber \\
&& \times \left[ u_{i\downarrow}^\dagger(1') u_{j\downarrow}^\dagger(2') d_{k\uparrow}^\dagger(3')+
u_{i\downarrow}^\dagger(1')d_{j\downarrow}^\dagger(2')u_{k\uparrow}^\dagger(3') +
d_{i\downarrow}^\dagger(1')u_{j\downarrow}^\dagger(2')u_{k\uparrow}^\dagger(3')
 \right]|0\rangle
\nonumber \\
&& ~~~ + k_{2\perp}^{\prime -} \psi^{(3)}_{\Delta} (1',2',3')  \\
&&\left. \times \left[ u_{i\downarrow}^\dagger(1')
u_{j\uparrow}^\dagger(2') d_{k\uparrow}^\dagger(3')+
u_{i\downarrow}^\dagger(1') d_{j\uparrow}^\dagger(2')
u_{k\uparrow}^\dagger(3')+ d_{i\downarrow}^\dagger(1')
u_{j\uparrow}^\dagger(2') u_{k\uparrow}^\dagger(3')
\right]|0\rangle \right\} + ...\nonumber
\end{eqnarray}
where $k_{\perp}^\pm =k_x \pm ik_y$, the ellipses denote other components of the wave functions which do not
enter the following calculation, and $\uparrow$ ($\downarrow$) on the quark creation operators denotes the
positive (negative) helicity of the quarks. The amplitudes $\psi^{(1)}_{P,\Delta}$ ($\psi^{(3)}_{P,\Delta}$,
$\psi^{(4)}_P$) have zero (one) unit of the orbital angular momentum projection. $\psi^{(1)}_\Delta(1,2,3)$ is
symmetric in 1 and 2.  The integration measure is given by:
\begin{eqnarray}
d[1]d[2]d[3] &=&  \frac{dx_1dx_2dx_3}{\sqrt{x_1 x_2 x_3}}
\frac{d^2k_{1\perp}d^2k_{2\perp}d^2k_{3\perp}}{(2\pi)^9}
\nonumber\\
&&\  \times 2\pi \delta(1-x_1-x_2-x_3)(2\pi)^2 \delta^{(2)}(\bf
k_{1\perp}+\bf k_{2\perp}+\bf k_{3\perp})
\end{eqnarray}
and the $x_i$ are the fractions of the proton linear momentum $P_N$ carried by the quarks, and $k_{i\perp}$ are
the corresponding transverse momenta. For the $\Delta$ we  use $y_i$ instead of $x_i$ and $k_{i\perp}^{\prime}$
instead of $k_{i\perp}$. It should be noted that since the $\Delta$ moves in  $-z$ direction then the orbital
angular wave function must be $k_{2\perp}^{\prime -}$ as shown in Eq.\ (5). The matrix element $G_0$ has two
contributions: one in which the proton has $l_z=0$ and the $\Delta$ carries $l_z=-1$ and in the second one the
proton has $l_z=1$ and the $\Delta$ carries $l_z=0$. For the first case, we introduce the hard amplitudes
$T_i(1,2,3,1^\prime,2^\prime,3^\prime)$, which represent three quark scattering off the external photon with
two-gluon exchange. The index $i=1,2,3$ indicates that the photon is attached to $i$-th quark. For the second
case, the hard amplitudes are denoted as $T_i^{\prime}(1,2,3,1^\prime,2^\prime,3^\prime)$, in which
all quark helicities are reversed.
We will explain how to compute these hard amplitudes in the next
section.

\section{Leading-Order pQCD Factorization Formula for $G_0$}

In this section, we derive a leading-order pQCD factorization
formula for the helicity-flip $N\gamma^*\rightarrow \Delta$
matrix element $G_0$. As shown in Eq. (1), this is generated by a
virtual photon with longitudinal polarization, corresponding to
the Coulomb quadrupole amplitude.

The first step in calculating $G_0$ is to obtain the hard
scattering amplitudes of partons. For this we follow the
guidelines of Ref. \cite{beli}. In principle, one has to consider
both three-quark scattering with one unit of orbital angular
momentum or three-quark-one-gluon scattering without orbital
motion. Here we consider explicitly only the former and take into
account the latter through the requirement of color gauge
invariance, leaving out the contribution with the gluon field
strength tensor. This practice of leaving out dynamical gluon
contribution corresponds to the so-called Wendzura-Wilczek
approximation in the literature \cite{ww}. A key point in the
calculation is that since the quark masses are negligible, the
quark helicity is conserved. This is due to the vectorial nature
of the electromagnetic coupling between the struck quark and the
virtual photon and of QCD quark-gluon coupling \cite{lep}. When a
Fock component of the light-cone wave function of the proton
contains two quarks of the same flavor and helicity then,
clearly, it contributes only when there are two quarks of the
same flavor and helicity in the $\Delta$ wave function. The
exchange contribution that results due to the presence of the
same two quarks has to be taken into account properly through the
second quantized calculation.

\begin{figure}[t]
\begin{center}
\includegraphics[height=9.0 cm]{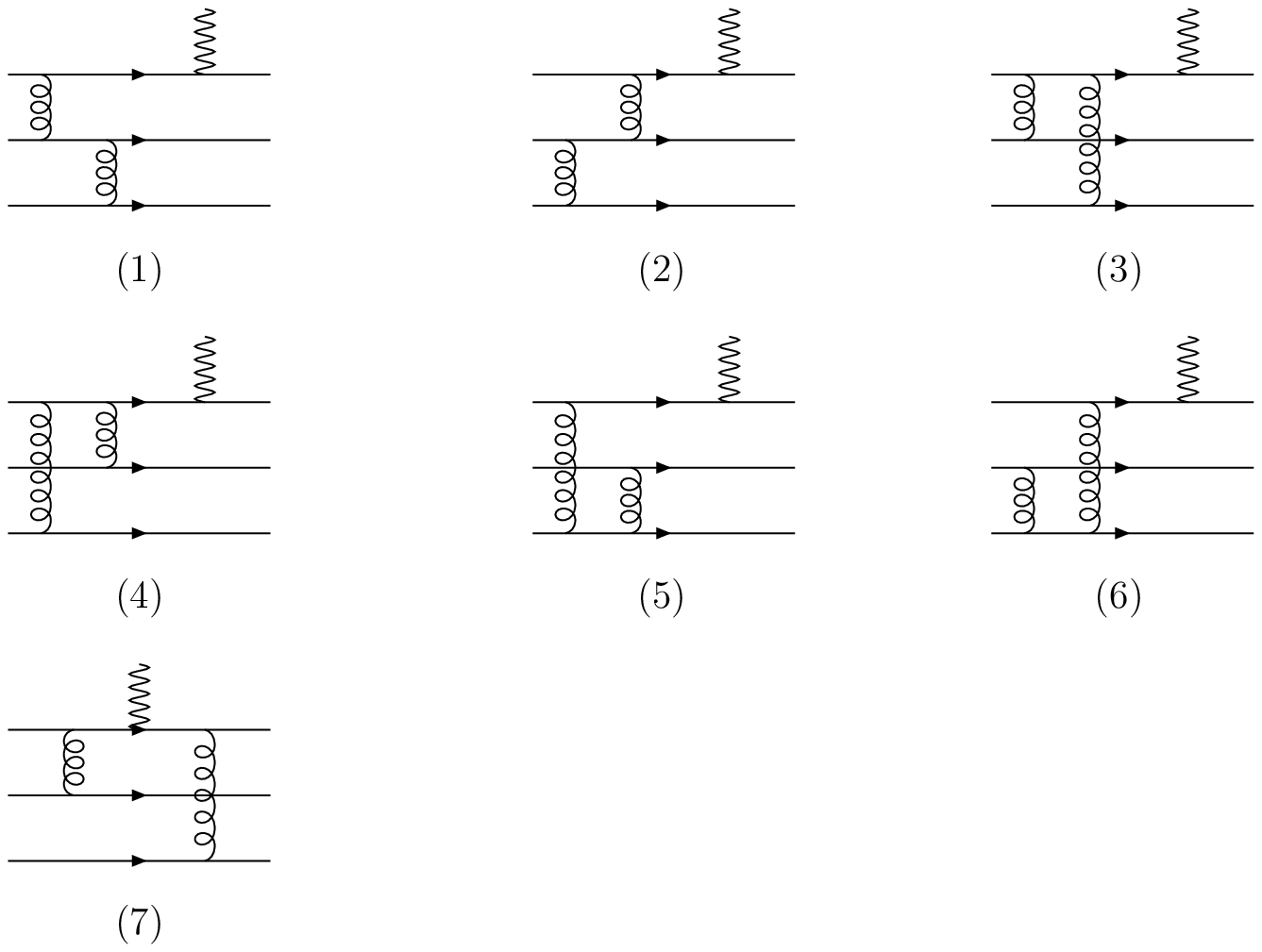}
\end{center}
\caption{The leading pQCD diagrams contributing to the nucleon to
$\Delta$ transition amplitudes. The mirror diagrams should be
added.}
\end{figure}

Using the light-cone wave functions of the proton and the $\Delta$
in Eqs.(4) and (5) and keeping terms that are linear in the quark
transverse momenta, one obtains:
\begin{eqnarray}
G_0 &=& -\frac{e_u-e_d}{24 (2M_N) } \int d[1][2][3]d[1^\prime ][2^\prime ][3^\prime ]
\left\{k'^-_{1\perp}\psi^{(3)*}_\Delta(2^\prime ,1^\prime ,3^\prime )
     \psi^{(1)}_P (1,2,3) \right.
\nonumber\\
 && \times \left[ T_2(1,2,3,1^\prime ,2^\prime ,3^\prime ) + T_2(1,2,3,3^\prime ,2^\prime ,1^\prime )
  - T_3(1,2,3,1^\prime ,2^\prime ,3^\prime ) -T_3(1,2,3,3^\prime ,2^\prime ,1^\prime ) \right]
   \nonumber\\
  &&  +\psi^{(1)*}_\Delta (1^\prime ,3^\prime ,2^\prime )
     (k_{2\perp}^+
           \psi_P^{(3)}
         + k_{1\perp}^+ \psi_P^{(4)}) (2,1,3)
 \nonumber\\
 && \left. \times \left[ T'_2(1,2,3,1^\prime ,2^\prime ,3^\prime )
         + T'_2(3,2,1,1^\prime ,2^\prime ,3^\prime )
         - T'_3(1,2,3,1^\prime ,2^\prime ,3^\prime )
         -T'_3(3,2,1,1^\prime ,2^\prime ,3^\prime ) \right]
  \right\}
\end{eqnarray}
where $T_i$ and $T_i'$ are the three-quark scattering amplitudes
introduced in the previous section. The color wave function of
the three quarks is normalized to unity.

The perturbative Feynman diagrams for $T_i$ are given in Fig.\ 1.
There are fourteen such diagrams, taking into account the
relative position of the photon vertex and the gluon interactions.
In each diagram we have three collinear incoming and outgoing
quarks exchanging two gluons. The photon could interact with any
one of the quarks. With the spin-isospin structure of the proton
and the $\Delta$ state functions, we assume that the first and
third quarks have positive helicities while the second quark has
negative one.

To find the amplitudes $T_i$, we let the outgoing quarks carry
orbital angular momenta $k^{\prime}_{i\perp}$ and the incoming
quarks have zero orbital angular momenta, {\it i.e.},
$k_{i\perp}=0$. From each one of the Feynman diagrams we write the
amplitude following the usual QED and QCD Feynman rules. Since the
calculation is performed in the collinear frame in which the
particles are highly relativistic, we can set the masses of the
proton and the $\Delta$ to zero. We expand the quark spinors in
the final state and the quark propagators to first order in the
transverse momenta. Then we collect all such contributions from
the given diagram and sum up the results of all the Feynman
diagrams. This yields $T_i(k^{\prime}_{i\perp},x_i,y_i,Q^2)$. To
find $T^{\prime}_i(k_{i\perp},x_i,y_i,Q^2)$ we set
$k^{\prime}_{i\perp}=0$ and $k_{i\perp} \neq 0$ and follow similar
steps.

The results for $T_1(1,2,3,1',2',3')$ at order of ${\cal O}({\bf
k_\perp})$ are as follows:
        \begin{eqnarray}
         T_1(1,2,3,1',2',3') &=& -g_s^4 \frac{8 C_B^2}{ Q^4} \left[k'^+ _{1\perp} w_1(x_1,x_2,x_3,y_1,y_2,y_3) +
     k'^+ _{3\perp} w_2(x_1,x_2,x_3,y_1,y_2,y_3) \right],
     \end{eqnarray}
where $C_B=2/3$ and the functions $w_1$ and $w_2$ are:
    \begin{eqnarray}
     w_1(x_1,x_2,x_3,y_1,y_2,y_3) &=& \frac{1}{\bar{x}_1\bar{y}_1^2 x_3^2y_3}
     +\frac{1}{\bar{x}_1^2\bar{y}_1^2 x_3y_3}
        -\frac{1}{\bar{y}_1\bar{x}_3x_2y_2x_3y_3} +\frac{1}{\bar{y}_1^2\bar{x}_1^2 x_2y_2}
    \nonumber\\
    && -\frac{1}{\bar{x}_3 y_1y_2y_3x_2x_3^2}
    \nonumber\\
    w_2(x_1,x_2,x_3,y_1,y_2,y_3) &=& \frac{1}{\bar{x}_1\bar{y}_1 x_3^2y_3^2}
     +\frac{1}{\bar{x}_1^2\bar{y}_1 x_3 y_3^2}
      -\frac{1}{{\bar x}_3y_2 x_2x_3y_3^2}
       \nonumber\\
            && + \frac{1}{\bar{x}_1^2\bar{y}_1y_3 x_2y_2}
        + \frac{1}{\bar{x}_3 x_2 y_2x_3^2y_3^2}.
\end{eqnarray}
Similarly for $T_2$ we have:
\begin{eqnarray}
 T_2(1,2,3,1',2',3') &=& -g_s^4 \frac{8C_B^2}{Q^4} \left[ k^{\prime +} _{1\perp} w_0(y_1,y_2,y_3,x_1,x_2,x_3) +
     k^{\prime +} _{3\perp} w_0(y_3,y_2,y_1,x_3,x_2,x_1) \right],
     \end{eqnarray}
    the function $w_0$ is given by:
    \begin{eqnarray}
    w_0(x_1,x_2,x_3,y_1,y_2,y_3) &=&-\frac{1}{\bar{x}_2\bar{y}_2 x_1 x_3 y_3^2}
        +\frac{1}{ x_1^2 y_1^2 x_3 y_3} +\frac{1}{\bar{x}_1 x_1y_1x_3y_3^2}
        \nonumber\\
         && + \frac{1}{\bar{x}_2\bar{y}_2 x_1^2 y_1^2}
         +\frac{1}{\bar{x}_1 \bar{y}_3 y_1x_1^2 x_3y_3}.
         \end{eqnarray}

         To obtain $T_3(1,2,3,,1',2',3')$ we interchange
         $x_1,y_1,k^{\prime +}_{1\perp}$ and $x_3,y_3,k^{\prime
         +}_{3\perp}$ and to obtain $T'_i(1,2,3,1',2',3')$ we
         replace $k^{\prime +}_{i\perp}$ with $k^{+}_{i\perp}$ and interchange
         $x_i$ and $y_i$. The above results are consistent with
         those derived in Ref. \cite{beli}.

To further simplify the above expressions, let us define
$k_\perp$-intergrated wave functions:
         \begin{eqnarray}
            \phi^{(3)}_ {P}(x_1,x_2,x_3) &=& 2 \int
            [d^2{\bf k_\perp} ] \psi^{(1)}_P (k_1,k_2,k_3)
            \nonumber \\
            \phi^{(4)}_P(x_1,x_2,x_3) &=& 2 \int [d^2{\bf k_\perp} ]
            {\bf k_{3 \perp}}\cdot ({\bf k_{2\perp}}\psi^{(3)}_P
             +{\bf k_{1\perp}}\psi^{(4)}_P) (k_2,k_1,k_3),
             \nonumber\\
             \psi^{(4)}_P(x_1,x_2,x_3) &=& 2 \int [d^2{\bf k_\perp} ]
               {\bf k_{1 \perp}}\cdot ({\bf k_{2\perp}}\psi^{(3)}_P
               +{\bf k_{1\perp}}\psi^{(4)}_P)
               (k_2,k_1,k_3) \nonumber \\
                       \phi^{(3)}_{\Delta}(x_1,x_2,x_3) &=& 2 \int [d^2{\bf k_\perp} ] \psi^{(1)}_{\Delta}
  (k_1,k_2,k_3) \nonumber \\
         \phi^{(4)}_{\Delta}(x_1,x_2,x_3) &=& 2 \int [d^2{\bf
            k_\perp}] {\bf k_{3\perp}} \cdot {\bf k_{1\perp}}
            \psi^{(3)}_\Delta(k_2,k_1,k_3),\nonumber \\
            \psi^{(4)}_{\Delta}(x_1,x_2,x_3) &=& 2 \int [d^2{\bf
            k_\perp} ] {\bf k_{1\perp}} \cdot {\bf k_{1\perp}}
            \psi^{(3)}_\Delta (k_2,k_1,k_3)
             \end{eqnarray}
               where the integration measure is defined as:
               \begin{eqnarray*}
                [d^2{\bf k_\perp}] &=& \frac {d^2 k_{1\perp}
                d^2k_{2\perp} d^2k_{3\perp}}{(2\pi)^6}
                \delta^{(2)}(\bf k_{1\perp}+ \bf k_{2\perp}+ \bf
                k_{3\perp}).
                \end{eqnarray*}

               In terms of these wave functions and by using the
               obtained expressions for the hard amplitudes we get
               the following expression for $G_0$:
        \begin{eqnarray}
               G_0 &=& (4\pi\alpha_s)^2 \frac{C_B^2}{ (2M_N)  4Q^4}\frac{e_u-e_d}{3}\int [dx][dy]
               \nonumber\\
                && \Big\{ \phi^{*(4)}_{\Delta}(y_1,y_2,y_3)\phi^{(3)}_P(x_1,x_2,x_3)
                 \Big[ w_0(y_1,y_2,y_3,x_1,x_2,x_3)+w_0(y_1,y_2,y_3,x_3,x_2,x_1)
                 \nonumber\\
                  && -w_1(x_3,x_2,x_1,y_1,y_2,y_3) -w_2(x_3,x_2,x_1,y_3,y_2,y_1) \Big]
                  \nonumber\\
                  && +\psi^{*(4)}_{\Delta}(y_1,y_2,y_3)\phi^{(3)}_P(x_1,x_2,x_3)
                  \Big[ w_0(y_3,y_2,y_1,x_3,x_2,x_1) + w_0(y_3,y_2,y_1,x_1,x_2,x_3)
                  \nonumber\\
                   && -w_1(x_3,x_2,x_1,y_3,y_2,y_1) -w_2(x_3,x_2,x_1,y_1,y_2,y_3) \Big]
                    \nonumber\\
                    && +\phi^{(4)}_P(x_1,x_2,x_3)\phi^{*(3)}_\Delta (y_1,y_3,y_2)
             \Big[ w_0(x_3,x_2,x_1,y_3,y_2,y_1) +w_0(x_3,x_2,x_1,y_1,y_2,y_3)
             \nonumber\\
             && -w_1(y_3,y_2,y_1,x_3,x_2,x_1) -w_2(y_3,y_2,y_1,x_1,x_2,x_3)\Big]
             \nonumber\\
             && +\psi^{(4)}_P(x_1,x_2,x_3)\phi^{*(3)}_\Delta (y_1,y_3,y_2)
    \Big[ w_0(x_1,x_2,x_3,y_1,y_2,y_3) + w_0(x_1,x_2,x_3,y_3,y_2,y_1)
    \nonumber\\
    &&  -w_1(y_3,y_2,y_1,x_1,x_2,x_3)-w_2(y_3,y_2,y_1,x_3,x_2,x_1) \Big]
    \Big\},
    \end{eqnarray}
    where $[dx]= dx_1 dx_2 dx_3 \delta (1-x_1-x_2-x_3)$ and
    $\alpha_s = g_s^2/4 \pi$.
    We see from Eq.(13) that $G_0$ scales like $1/Q^4$ in the high
    $Q^2$ limit, consistent with the general power counting. To find the normalization of $G_0$ we need to know the
    light-cone distribution amplitudes defined in Eq.(12). For the
    proton, a set of such functions have been given in Ref.
    \cite{bra} based on conformal expansion, QCD sum rules and Lorentz
symmetry. For the $\Delta$ no such work has been done
    explicitly. However if the $\Delta$ is to be treated as a
    three-quark state, we believe that, at the asymptotically
    large $Q^2$, the light-cone distribution amplitudes of the
    $\Delta$ will have a $y_i$-dependence which is identical to
    the $x_i$-dependence of the light-cone distribution amplitudes
    of the proton \cite{future}. For the proton we have, for example, the
    asymptotic form of $\phi_p^{(3)}$ is $x_1 x_2 x_3$ and
    $\phi_p^{(4)}$ is $x_1 x_2$ \cite{brod1}. If we perform the
    $x_i$ and $y_i$ integration in Eq.(13) with such functions we
    get, due to end-point singularities, a double logarithmic
    divergence results. This divergence indicates the quarks with small
    Feynman $x$ contribute significantly to the hard scattering.
    However, for very small $x$ for which the parton longitudinal
    momentum is on the order of $\Lambda_{\rm QCD}$, the hard
    scattering picture breaks down, and the above calculation is
    invalid. Then one has to add the so-called Feynman contribution
    which is relatively unimportant in the limit of large $Q^2$.

    Therefore, these divergent integrals are regulated physically by
    a cut-off from below of order $\Lambda^2/Q^2$ where
    $\Lambda$ is some parameter that represents the soft energy
    scale (order of few hundreds of MeV).
    With this cut-off, the result of the momentum fraction
    integration will be a $Q^2$-dependent term of the form,
    $\log^2{(Q^2/\Lambda^2)}$. Thus we can write:
    \begin{equation}
    G_0 = c^{\prime} \log^2{(Q^2/\Lambda^2)}/Q^4
    \end{equation}
    where $c^{\prime}$ is a numerical factor that depends on the
    explicit expressions of the light-cone distribution
    amplitudes. With this form of $G_0$ and with the fact that $G_-$ is of
    order $1/Q^2$ relative to $G_+$ we will give in the next
    section a phenomenological prediction of the ratio $R_{SM}$ in the high $Q^2$
    limit.

    \section{Phenomenology}

    In the electro-production  of the $\Delta$ resonance, the multipoles $M1$, $E2$ and $C2$ of
    the exchanged virtual photon are the only ones that
    contribute \cite{scad,dur, duf,dev}. One of the experimentally extracted quantities related
    to this process is the ratio $R_{SM} \equiv C2/M1$. In order
    to express this ratio in terms of the helicity matrix elements $G_{\lambda}$ we introduce
    the resonance ``helicity amplitudes'' $A_{1/2}$, $A_{3/2}$ and
    $S_{1/2}$ \cite{cop}. These amplitudes are computed in the rest frame of the $\Delta$ and the sub-indices
    refer to the helicity of the $\Delta$. The scalar amplitude $S_{1/2}$ is relevant only
     for virtual photons. These amplitudes are related to the helicity matrix elements
     through the relations \cite{sto}:
    \begin{equation}
    A_{1/2}= \eta G_+ \; ; A_{3/2}  = \eta G_- \; ; S_{1/2}= \eta \frac {\vert \bf q
    \vert}{Q}G_0
    \end{equation}
    where $\eta$ is some kinematical factor and $\vert \bf q
    \vert$ is the photon three-momentum in the rest frame of the
    $\Delta$ given by : $\vert {\bf q} \vert =\sqrt{ \left ( \frac
    {M^2_\Delta+M^2_N+Q^2}{2M_\Delta}\right ) ^2-M_N^2}$.
    The multipole $C2$ is proportional to $S_{1/2}$ and we also
    have \cite{wit}:
    \begin{equation}
    M1= - \frac {1}{2} A_{1/2} - \frac{\sqrt 3}{2} A_{3/2}
    \end{equation}
    From the results of the previous section we see that $A_{3/2}$
    is of order $1/Q^2$ relative to $A_{1/2}$ in the high $Q^2$
    limit and thus it will be neglected. With this approximation the
    ratio $R_{SM}$ becomes:
     \begin{equation}
    R_{SM}= \frac{\vert \bf q \vert}{Q}\frac{G_0}{G_+}
    \end{equation}
     Substituting in the last equation the expression for $ \vert \bf q \vert$ and using
    Eq.(14) we get:
     \begin{equation}
    R_{SM}=c \  \sqrt {\left (\frac {2.4+Q^2}{2.464}\right )^2-.88} \  \frac {\log^2
    {(Q^2/\Lambda^2)}}{Q^2}
    \end{equation}
    The pre-factor $c$ is determined by taking $\Lambda = .25$ GeV and  by fitting Eq. (18) with the
    recently obtained experimental values of $R_{SM}$ \cite{fro}:
    \begin{eqnarray*}
    R_{SM} &=& -.112 \pm .013\  \rm {at}\  Q^2= 2.8\ GeV^2,\\
     R_{SM} &=&-.148 \pm .013\ \rm {at}\ Q^2= 4.0\ GeV^2,
    \end{eqnarray*}
    The result is:
     \begin{equation}
    R_{SM}=- \frac {.013 \  \sqrt {-.88+(2.4+Q^2)^2} \  \log^2 {(16
    Q^2)}}{Q^2}
    \end{equation}

    \begin{figure}[t]
\begin{center}
\includegraphics[height=9.0 cm]{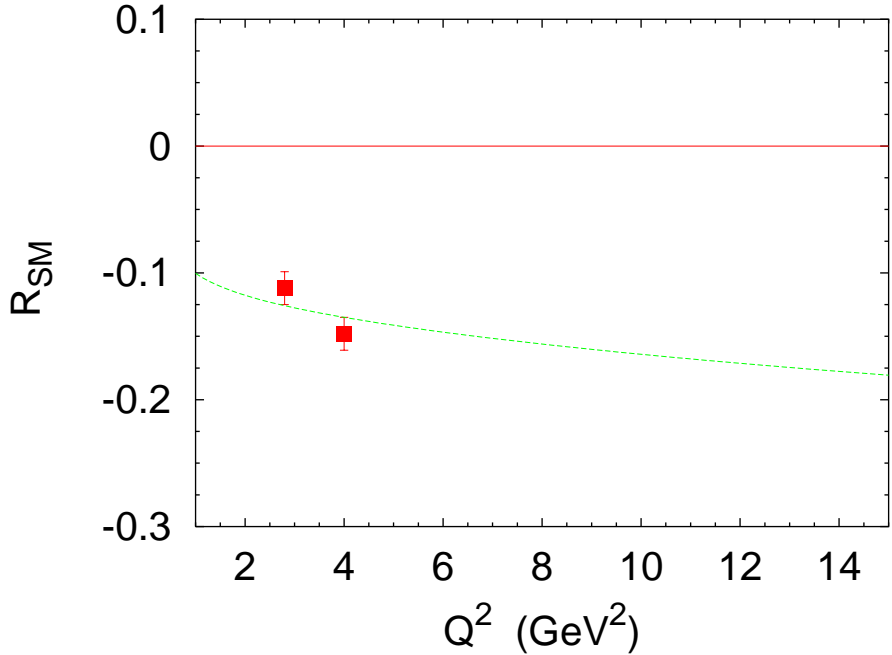}
\end{center}
\caption{A phenomenological prediction for the ratio $R_{SM}$.}
\end{figure}
$R_{SM}$ is displayed in Figure 2. From that figure it is clear that $R_{SM}$ has a slow variation at the
presently-accessible momentum transfer $Q^2$ at JLab which is about 6 $\sim$ 7 GeV$^2$. We remark here that for
the E2-to-M1 ratio $R_{EM}$, two units of orbital angular momentum are need to induce the transition, and higher
powers of the logarithmic correction may arise when one tries to calculate the matrix element $G_-$.

\section{Summary}

A perturbative QCD calculation of the helicity flip matrix element of the electromagnetic $N \rightarrow \Delta$
transition has been given. We showed that the transverse momenta of the quarks in the proton and $\Delta$ is
essential to obtain a finite result. An explicit calculation of the hard amplitude $T_i(x_i,y_i,
k^{\prime}_{i\perp}, Q^2)$ and $T^{\prime}_i(x_i,y_i, k_{i\perp}, Q^2)$ has been given and the techniques of the
calculation were outlined in some detail. The essential steps of the calculation are to draw the relevant
perturbative Feynman diagrams and then to compute the contribution from each such diagram by expanding the quark
spinors and propagators to first power of the transverse momenta. The hard amplitudes are then to be convoluted
with light-cone amplitudes. Our pQCD results for the scaling behavior of $G_+$, $G_0$ and $G_-$ confirm earlier
scaling prediction \cite{carl}. In addition we outlined how one obtains the double logarithmic correction. Based
on our result for $G_{\lambda}$ we gave a phenomenological prediction for the ratio $R_{SM}$. Roughly speaking,
$\vert R_{SM} \vert$ will be of order of $20 \% $ at $Q^2$ of order 10 GeV$^2$.

We thank A.V. Belitsky, C. Carlson, P. Stoler, and F. Yuan for a
number of useful discussions.

\end{document}